# Decoupled nematic and magnetic criticality in FeSe$_{1-x}$S$_x$

Jake Ayres,[1,2]* Matija Čulo,[2] Jonathan Buhot,[1,2] Bence Bernáth,[2] Shigeru Kasahara,[3] Yuji Matsuda,[3] Takasada Shibauchi,[4] Antony Carrington,[1] Sven Friedemann,[1] Nigel E. Hussey[1,2]†

[1] H. H. Wills Physics Laboratory, University of Bristol, Tyndall Avenue, Bristol BS8 1TL, United Kingdom.
[2] High Field Magnet Laboratory (HFML-EMFL) and Institute for Molecules and Materials, Radboud University, Toernooiveld 7, 6525 ED Nijmegen, Netherlands.
[3] Department of Physics, Kyoto University, Sakyo-ku, Kyoto 606-8502, Japan.
[4] Department of Advanced Materials Science, University of Tokyo, Kashiwa, Chiba 277-8561, Japan.

* jake.ayres@bristol.ac.uk
† n.e.hussey@bristol.ac.uk

Electronic nematicity in correlated metals often occurs alongside another instability such as magnetism. As a result, the question remains whether nematicity alone can drive unconventional superconductivity or anomalous (quantum critical) transport in such systems. In FeSe, nematicity emerges in isolation, providing a unique opportunity to address this question. Studies to date, however, have proved inconclusive; while signatures of nematic criticality are observed upon sulfur substitution, they appear to be quenched under the application of pressure due to the emergent magnetism. Here, we study the temperature and pressure dependence of the low-temperature resistivity of FeSe$_{1-x}$S$_x$ crystals at $x$ values just beyond the nematic quantum critical point. Two distinct components to the resistivity are revealed; one whose magnitude falls with increasing pressure and one which grows upon approaching the magnetic state at higher pressures. These findings indicate that nematic and magnetic critical fluctuations in FeSe$_{1-x}$S$_x$ are completely decoupled, in marked contrast to other Fe-based superconductors, and that nematic fluctuations alone may be responsible for the transport signatures of quantum criticality found in FeSe$_{1-x}$S$_x$ at ambient pressure.

**Introduction**

A common characteristic of unconventional superconductors is their proximity to another ground state of broken symmetry, fluctuations of which can both mediate superconductivity and drive non-Fermi-liquid (nFL) behavior in the vicinity of its associated quantum critical (QC) point. Nematicity – a lowering of rotational symmetry without breaking translational symmetry – is one form of order that has been observed in a variety of systems, including iron-based (*1-3*), cuprate (*4*), heavy fermion (*5*) and Moiré (*6*) superconductors. The extent to which nematic order and its fluctuations are responsible for pairing and QC phenomena has proved a challenging question, however, largely due to the fact that nematicity often occurs in the vicinity of another, possible primary, instability. In iron pnictides, for example, nematicity is claimed to be a spin-driven effect (*7*) while QC phenomena observed in Sr$_3$Ru$_2$O$_7$ – initially attributed to a nematic quantum critical point (NQCP) (*8*) – were later found to arise in the presence of a field-tuned spin-density wave (*9*).

FeSe is unusual in that nematic order stabilizes in the absence of static magnetism (*3*). Below a tetragonal-to-orthorhombic distortion at $T_s$ = 90 K, both its normal (*10*) and superconducting (SC) (*11*) state properties exhibit marked two-fold anisotropy. Although widely believed to be electronic in origin (*12*), it remains unclear whether the nematic transition is driven by charge (*13*), orbital (*14*) or magnetic (*15*) correlations. Nevertheless, its discovery offers a unique opportunity to test theoretical predictions for nFL or 'strange metallic' behavior arising solely from critical nematic fluctuations (*16-22*). To this end, a large effort has been made to elucidate the respective roles of nematic and magnetic fluctuations in shaping the normal and SC properties of FeSe (*23-26*).

High-pressure studies on FeSe have proved to be highly instructive in this pursuit. As pressure increases, $T_s$ is suppressed (to $T_s$ = 0 K at $p = p_c$) but the SC transition temperature $T_c$ is not enhanced at $p_c$ (*27*). Beyond the nematic state ($p > p_c$), however, there is a marked (four-fold) increase in $T_c$ (*27-29*) that has been naturally linked to strengthening magnetic interactions (*30*). The role of nematicity in driving nFL/QC phenomena has proved more controversial. At $p = p_c$, the critical nematic fluctuations in FeSe are quenched (*13*), presumably due to the emergence of long-range magnetic order before the nematic phase terminates (*31*). In FeSe$_{1-x}$S$_x$, nematicity is also suppressed with increasing sulfur substitution, vanishing at a critical S concentration $x_c$ = 0.17 (*32*) where the nematic susceptibility also diverges (*12*). Since no magnetic order develops at any point across the substitution series (at ambient pressure), this divergence suggests that a genuine NQCP exists in FeSe$_{1-x}$S$_x$.

The question remains, however, whether the emergent critical nematic fluctuations are responsible for the strange metal transport seen at ambient pressure in FeSe$_{1-x}$S$_x$ (*33-38*). Low energy spin-fluctuations, for example, are known to persist to low $T$ and low $x$ (*39*). Moreover, the lack of divergence in the effective mass $m^*$ on approaching the NQCP (*40*) has been viewed as evidence that the critical nematic fluctuations are also quenched at ambient pressure, in this case due to nemato-elastic coupling or local strain effects (*41,42*), the nFL transport then being attributed to scattering off the residual spin fluctuations.

With increasing $x$, $p_c$ falls while $p_m$, the onset pressure for magnetic order, increases (*43*), leading ultimately to a separation of the nematic and magnetic phases in the ($p$, $T$) plane at higher $x$. Previous NMR measurements appeared to confirm such a separation at $x = 0.12$ (< $x_c$) (*44*). Detailed transport studies on pressurized FeSe$_{1-x}$S$_x$ with $x = 0.11$ (*45*) then revealed the absence of nFL transport or $m^*$ enhancement across $p_c$, supporting the picture of quenched nematic criticality due to strong nemato-elastic coupling (*45*). A more recent $\mu$SR study, however, found that magnetism at $x = 0.12$ is stabilized *before* nematicity is destroyed (the discrepancy between $\mu$SR and NMR likely reflecting the different timescales of the two probes). Hence, it is unclear whether the suppression of nematic criticality near $x = 0.12$ under pressure is due to coupling to the lattice or to slowly fluctuating moments. In order to determine whether critical nematic fluctuations alone can drive nFL transport in FeSe$_{1-x}$S$_x$, pressure studies on samples with higher $x$ values, where the nematic and magnetic phases are fully separated, are required.

Here, we study the low-$T$ resistivity $\rho(T)$ of FeSe$_{1-x}$S$_x$ with $x = 0.18$ and 0.20 (> $x_c$) under applied pressures up to 15 kbar (< $p_m$). Whilst the form of $\rho(T)$ cannot differentiate easily between nematic and magnetic fluctuations, tracking its evolution with $p$ may reveal an approach to or a retreat from a QCP associated with either order parameter. In this way, their respective influences can be disentangled. For both samples studied here, we find two distinct $T^2$ components in $\rho(T)$ (due to quasiparticle-quasiparticle scattering) which extend over different $T$ ranges and whose coefficients show contrasting $p$-dependencies. The term that grows with increasing $p$ is attributed to the dressing of quasiparticles by critical magnetic fluctuations that strengthen upon approach to the magnetic QCP (*43,46*). Its coefficient at ambient pressure, however, is found to be negligible. This result provides strong evidence that the source of the large and strongly $x$-dependent $T^2$ coefficient observed at ambient pressure is scattering of quasiparticles that are dressed purely by the orbital nematic fluctuations. Finally, this coexistence of two distinct components to $\rho(T)$ also implies that, in contrast to what is observed in the iron-pnictides, the critical nematic and magnetic fluctuations in FeSe$_{1-x}$S$_x$ are completely decoupled.

**Results and Discussion**

Figure 1A and 1B show, respectively, the zero-field $\rho(0, T)$ (pale) and high-field $\rho(35\text{ T}, T)$ curves for samples with nominal $x$ values of 0.18 and 0.20 oriented **H**//**I**//$ab$ at various pressures $0 \leq p \leq 14.4$ kbar. The suppression of superconductivity by the magnetic field is apparent in all data sets. For $T > T_c$, there is almost complete overlap between $\rho(0, T)$ and $\rho(35\text{ T}, T)$, confirming that the magnetoresistance in this field orientation is negligible beyond $x_c$ (*34,36*), in marked contrast to the large magnetoresistance seen for **H**//$c$ (*35,36*). The broadening and structure of the superconducting transitions in $\rho(0, T)$ is highly reproducible between subsequent cooldowns at different pressures, between samples of similar dopings (*47*) and between measurements performed by different groups (*46,48*) indicating that non-hydrostaticity is unlikely to be playing a role here. We also note that the transitions sharpen again at higher pressures (~3 GPa) (*43*) suggesting that this behavior is in fact intrinsic.

The corresponding derivatives d$\rho$/d$T$(35 T) of the high-field curves, shown in panels C and D of Fig. 1, reveal a systematic evolution of $\rho(T)$ under applied pressure. To better orientate our discussion, we focus initially on the form of d$\rho$/d$T$ at ambient pressure. For $T < 10$ K, $\rho(35\text{ T}, T) = \rho_0 + A_tT^2$ with $A_t$ coefficients that are determined by fitting the d$\rho$/d$T$ traces below 10 K to a straight line through the origin (black lines in Fig 1C and 1D). We argue below that $A_t$ reflects the total quasiparticle-quasiparticle scattering cross-section enhanced by both magnetic and nematic critical fluctuations. Above the $T^2$ regime, d$\rho$/d$T$ is essentially flat, implying that $\rho(T)$ becomes $T$-linear (with coefficient $B$). Such a $T^2$ to $T$-linear crossover is characteristic of a metallic system in the vicinity of a QCP (*1, 49-52*).

A notable change in the derivative plots with increasing $p$ is the emergence of a finite linear slope in d$\rho$/d$T$ at higher temperatures, indicative of a second $T^2$ component that (i) coexists with the $T$-linear term, (ii) has a coefficient $A'$ that is around one order of magnitude smaller than $A_t$ and (iii) extends over a much broader temperature range. $A'$ and $B$ are determined by fitting the d$\rho$/d$T$ data between 20 K and 40 K to another straight line (black lines in Fig 1C and 1D). Whilst this second $T^2$ component is most evident in the derivative data at high $T$, the expectation is, as for a correlated Fermi liquid, that it extends down to lowest temperatures. In this way, $A_t$ is most naturally interpreted as the sum of two $T^2$ components, i.e. $A_t = A + A'$; the first component persisting up to ~ 10 K, the second component up to the highest temperature measured in our study (~ 40 K).

The $p$-dependence of coefficients $A$ ($A_t$), $B$ and $\rho_0$ (the latter obtained by extrapolating fits of the low-$T$ $\rho(T)$ curves at 35 T to 0 K) is shown in Figure 2A, B and C, respectively. It is immediately apparent that the relative slopes of all three quantities are the same, indicating that their $p$-dependencies share a common origin. The drop in all three with increasing pressure could signify either a reduction in scattering or an increase in the plasma frequency $\omega_p^2$ (i.e. $n/m^*$), or some combination thereof.

In the first scenario, the fall in $A$ ($A_t$), $B$ and $\rho_0$ with increasing $p$ (depicted in Fig. 2) would be attributed directly to a reduction in the dressing of quasiparticles by the relevant critical fluctuations. While this interpretation can support a typical quantum critical scenario in which $A(p)$ (and perhaps $\rho_0$) drops as the system is tuned away from the NQCP, the scattering rate associated with the linear-in-$T$ coefficient is not expected to decrease too (as inferred from Fig. 3B). Indeed, the $T$-linear resistivity inside of the quantum critical fan in FeSe$_{1-x}$S$_x$ at ambient pressure has been shown to be governed by a doping-independent scattering rate $1/\tau$ that is tied to the Planckian limit, i.e. $\hbar/\tau = ak_BT$ with $1 \leq a \leq 2$ (*34*).

In the second scenario, the change in all three coefficients can be ascribed wholly to an increase in $n/m^*$. Indeed, a sizeable increase in $n$ with pressure has been deduced in both FeSe (*51*) and FeSe$_{0.89}$S$_{0.11}$ (*45*) from quantum oscillation studies. Fig. 3A (Fig. 3B) shows the $p$-dependence of $A^*$, $B^*$ ($A'^*$), the coefficients $A$, $B$ and $A'$ rescaled by multiplying each quantity by $\rho_0(0)/\rho_0(p)$ (dashed lines in Fig. 2C) assuming the decrease in $\rho_0$ reflects a change in carrier density (and not a reduction in enhancement from the NQCP). As can be seen, the resultant $A^*$ and $B^*$ coefficients are either $p$-independent (for $x = 0.18$) or fall slightly (for $x = 0.20$) (note, however, the large error bars for the data at highest pressures). The near-constancy and magnitude of $B^*(p)$ is then consistent with the notion that the effective scattering rate remains at the Planckian bound with increasing pressure, in agreement with what had been found at ambient pressures (*34*). Within a QC scenario, the near-constancy of $A^*(p)$ is also consistent with the fact that the extent of the (low-$T$) $T^2$ regime in both samples does not vary with $p$. By contrast, at ambient pressure $A_t$ exhibits a marked decrease with increasing $x$ beyond the NQCP (see Fig. 3D) while the temperature of the $T^2$ to $T$-linear crossover increases as the system is tuned away from the NQCP by chemical substitution (*34,36*).

Irrespective of which scenario is the most appropriate, however, the clear increase in $A'$ (or in $A'^*$) with pressure, in both samples, is a robust observation. The order of magnitude change in $A'^*$, in particular, is even greater than that seen in $A_t^*$ upon approach to the NQCP at ambient pressure (Fig. 3D) and comparable to that observed in other quantum critical systems with well-established magnetic QCPs (*49, 53, 54*). Moreover, the fact that $A'$ is anti-correlated with $A$ and $B$ implies that the former has a distinct origin. We attribute the marked rise in $A'$ to an enhancement in the quasiparticle-quasiparticle scattering cross-section upon approach to a second, distinct QCP. The absolute magnitude of $A'$ over our experimental pressure range (~ 5 n$\Omega$cm/K$^2$), however, is much smaller than the value that $A_t$ reaches (> 200n$\Omega$cm/K$^2$ (*34,36*)) upon approaching the ambient pressure NQCP, as shown in Fig. 3D. This, coupled with the more extended temperature range over which this $T^2$ term persists, suggests that the second QCP may be situated at a critical pressure far beyond those accessible here. As illustrated in Fig. 2E, the approach to the second QCP also coincides with a marked (factor of 2) growth in $T_c$ for both samples, the growth in $A'$ and $T_c$ being largest for $x = 0.18$. As mentioned in the introduction, a marked increase in $T_c$ with pressure at lower sulfur concentrations has been linked previously to strengthening magnetic interaction (*30*).

Indeed, it has been suggested that $T_c$ is maximized at the magnetic QCP (*43*). Thus, it seems reasonable to associate this second QCP with the pressure-induced antiferromagnetic phase and to ascribe the *p*-dependence of the second $T^2$ component in $\rho(T)$ to quasiparticle-quasiparticle dressing by critical spin fluctuations in the quantum disordered regime. Of course, there are other scattering mechanisms that are capable of generating $T^2$ resistivity with a variable coefficient, such as non-critical electron-electron scattering near a Mott metal-insulator transition (*56*), electron-phonon scattering in disordered systems (*57*) or short-range spin fluctuation scattering (*58*). However, given the lack of evidence of Mottness in FeSe$_{1-x}$S$_x$ and the order of magnitude variation in *A′* over what is a relatively narrow range of applied pressures, these alternative explanations seem unlikely.

These contrasting *x*- and *p*-dependencies ($A_t(x)$ and $A^*(p)$) may be reconciled by considering the proposed $T = 0$ phase diagram shown schematically in Fig. 3C. The vertical solid- and open-headed arrows represent, respectively, the pressure tuning of the $x = 0.18$ and 0.20 samples, while the horizontal arrow represents tuning away from the NQCP with increasing *x* at ambient pressure. The near-constancy of *A\** (within the second scenario above) may indicate that $p_c(x)$ – the phase boundary for nematic order in the (*p*, *x*) plane – is very steep near $x = x_c$. This seems plausible given the steepness of $T_s(x)$ near $x_c$ – see Fig. 1A in Ref. (*38*), for example. Consequently, with increasing *p*, samples with $x > x_c$ track effectively parallel to the nematic phase boundary, rather than away from it. At the same time, the application of pressure tunes each sample towards $p_m(x)$ – the magnetic phase boundary – resulting in a marked increase in *A′*. In this way, the contrasting variation in *A(p)* and *A′(p)* can be understood. The steepness of the $p_c(x)$ boundary might also indicate a crossover in the nematic phase transition from second order to weakly first order near $x = x_c$. Such a crossover, intimated in Fig. 3C by the dashed nematic phase boundary, would lead to a cutoff in the nematic fluctuations, thereby providing an alternative explanation for the *p*-independence of *A\** and *B\**. It is noted that in pure FeSe, $T_s(p)$ terminates at a first-order structural and magnetic phase transition at ~ 2 GPa (a divergence of $1/T_1T$ at low *T* is lost) (*54*).

The presence of two anti-correlated but additive $T^2$ components in the low-*T* resistivity is unusual but implies the presence of two independent scattering channels of distinct origin. Given the correlation between *A′* and $T_c$ at finite pressure and the anti-correlation between *A′* and *A*, it seems very unlikely that spin fluctuations could be responsible for both. Indeed, while measurements of the spin-lattice relaxation rate in FeSe$_{1-x}$S$_x$ at ambient pressure indicate the emergence of low-lying spin fluctuations below $T_s$, spin fluctuations are strongly suppressed for $x > x_c$ (*39*). Moreover, as mentioned above, there is no evidence that such fluctuations go critical at $x = x_c$. It would appear that spin fluctuations, as parameterized by *A′* (~ $A_t/10$), play only a minor role in the overall low-*T* resistivity in FeSe$_{1-x}$S$_x$ at ambient pressure.

The measurements presented here imply that the nematic fluctuations anchored at the NQCP and the magnetic fluctuations anchored at the AFM QCP act as decoupled mechanisms for the enhancement of quasiparticle-quasiparticle scattering over most of the phase diagram of FeSe$_{1-x}$S$_x$. One possible way to account for their distinct nature is to consider the particular Fermi surface topology of FeSe$_{1-x}$S$_x$. Fig. 4A shows a schematic projection of the Fermi surface of FeSe$_{1-x}$S$_x$ ($x > x_c$) at $k_z = 0$ assuming only one hole pocket centered at Γ and two electron pockets at X and Y. Since spin fluctuations in detwinned FeSe are peaked at $\mathbf{Q} = (\pi, 0)$ (*15*), we also assume that in the tetragonal phase, critical spin fluctuations would enhance the quasiparticle-quasiparticle scattering cross-section predominantly at four 'hot-spots', as shown in Fig. 4B. The precise symmetry of the nematic fluctuations in FeSe$_{1-x}$S$_x$ has not yet been confirmed. Raman spectroscopy studies have indicated the presence of a *d*-wave Pomeranchuk instability (*58,59*) while quasiparticle scattering interference experiments (*60*) have revealed a highly anisotropic spectral weight (of different orbital character) on both pockets with *p*-wave symmetry (light shaded sections in Fig. 4B). For the former, critical nematic fluctuations would dress the quasiparticle states everywhere except at the AFM hot-spots (the nodes of the *d*-wave Pomeranchuk deformation), while for the latter, these cold-spots would reside at the 'bellies' of each pocket. These considerations might then help us to envisage how the influence of the critical nematic or magnetic fluctuations manifest themselves as two distinct components of the $T^2$ resistivity. Intriguingly, the in-plane magnetoresistance of FeSe$_{1-x}$S$_x$ (at ambient pressure) can also be decomposed into two components (*35*); a QC component that exhibits $H/T$ scaling and is maximal near the NQCP and a second component that remains purely $H^2$ (up to 35 T) and shows conventional Kohler's scaling. It is tempting to attribute these two components as arising from these distinct nematic and spin interactions, only one of which goes critical at ambient pressure.

Finally, we turn to consider the evolution of the superconductivity in FeSe$_{1-x}$S$_x$. While there is strong evidence to suggest that low-energy spin-fluctuations play a significant role in the pairing mechanism in FeSe$_{1-x}$S$_x$ (*23-26*), and the increase in $T_c$ ($p$) appears to be well correlated with $A'(p)$ (panels D and E of Fig. 2), it is striking that $A' \sim A_t/10$ at ambient pressure yet $T_c$ remains high (~ 8 K). This finding may suggest some role for nematicity in the pairing in FeSe$_{1-x}$S$_x$ but clearly, further work is required to confirm this. In pnictide superconductors, where nematicity and magnetism are strongly coupled, superconductivity is most likely driven by low-energy spin fluctuations, though $T_c$ could be enhanced by a reduction in the bare intra-pocket repulsion brought about by the nematic fluctuations (*7*). In the case of FeSe$_{1-x}$S$_x$, the decoupling of the nematic and the magnetic fluctuations means that this cooperative process is no longer viable and as a result, $T_c$ is not enhanced at the NQCP.

Previously, pressure tuning between two distinct QCPs was reported in the heavy fermion compounds Ge-doped CeCu$_2$Si$_2$ (*61,62*) and YbRh$_2$Si$_2$ with Ir and Co doping (*63*). To the best of our knowledge however, FeSe$_{1-x}$S$_x$ represents the first example of a correlated metal exhibiting an enhancement in the coefficient of the $T^2$ resistivity associated with two distinct QCPs. Clearly, the task is now to determine the universality classes associated with each criticality. In order to achieve this, however, it will be necessary to study a sample with a sulfur concentration even closer to the NQCP and to extend the pressure range (e.g. using an anvil cell) until the magnetic QCP itself is crossed. At the same time, determination of the evolution of complementary resistive properties (such as the Hall effect) with pressure may help elucidate further the nature of the two components.

**Materials and Methods**

Single crystals were grown via a KCl/AlCl$_3$ chemical vapour transport method. Their nominal dopings are $x = 0.18$ and 0.20. Both crystals were mounted together in a single piston-cylinder pressure cell and oriented such that **H** // *I* // *ab*. Daphne 7373, which is known to remain hydrostatic at room temperature up to 22 kbar (*65*), was used as a pressure transmitting medium. Resistivity measurements were performed using a standard four-point ac lock-in technique in Cell 4 of the High Field Magnet Laboratory (Radboud University, Nijmegen, The Netherlands) where a maximum magnetic field of 35 T could be applied. Temperature sweeps were performed in both field orientations (positive and negative 35 T) such that the longitudinal component could be isolated from any Hall component present due to an offset in the voltage contacts (though it is noted that the Hall contribution was found to be a near-negligible part of the total signal).

The actual S content of crystals can often be lower than the nominal value (12). For both of our samples, however, the zero-field $\rho(T)$ curves (at ambient pressure) are found to agree well with previous reports on samples with similar dopings (*33,36,37*). Specifically, there is no kink or minimum in the derivative $d\rho/dT$ that could be attributed to a finite $T_s$, and the $T^2$ regime at low-$T$ extends up to around 8-10 K with a coefficient $A_t \sim$ 40-55 n$\Omega$cm/K$^2$, compared with > 200 n$\Omega$cm/K$^2$ for $x \leq 0.17$ (*33,36*). While the as-measured $A_t$ values are ~ 25% lower than previous reports, this level of discrepancy is within the geometrical uncertainty associated with measuring small crystals inside a pressure cell.

**Acknowledgements**

The authors acknowledge enlightening discussions with M. Berben, C. Duffy, B. Goutéreaux, R. Hinlopen, Y.-T. Hsu and C. Pépin.

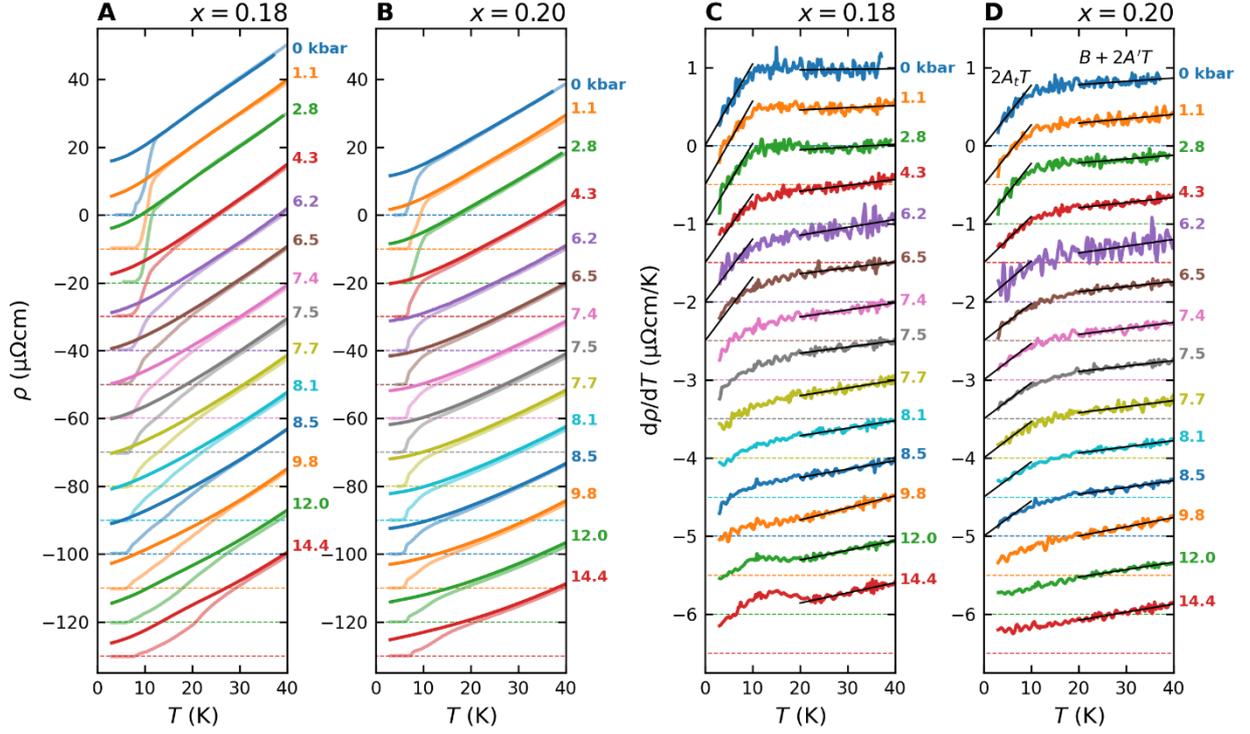

**Fig. 1. Pressure dependence of the high-field in-plane resistivities of FeSe$_{1-x}$S$_x$ beyond the NQCP. A (B):** Zero-field resistivity $\rho(0, T)$ (pale curves) and high-field resistivity $\rho(35$ T, $T)$ (dark curves) measured at indicated pressures between 0 and 14.4 kbar for $x = 0.18$ ($x = 0.20$) with **H**//*I*//*ab*. **C (D):** Corresponding first derivatives of the high-field resistivity for $x = 0.18$ ($x = 0.20$) at the same pressures as those in **A** (**B**). In all panels, an offset has been applied between subsequent pressures for clarity. The dashed lines indicate the $\rho = 0$ or d$\rho$/d$T$ = 0 position of each curve. In panels **C** and **D**, the black lines are straight-line fits below 10 K and above 20 K from which the resistivity coefficients $A_t$ ($A$), $B$ and $A'$ have been deduced (see text for details). The enhancement of superconductivity prevents $A_t$ from being determined at the highest applied pressures. The increasingly broad superconducting transitions manifest themselves as shallow peaks in the derivatives that are most visible in d$\rho$/d$T$($x = 0.18$) above 9.8 kbar (panel **C**) but may influence the data at lower pressures as well.

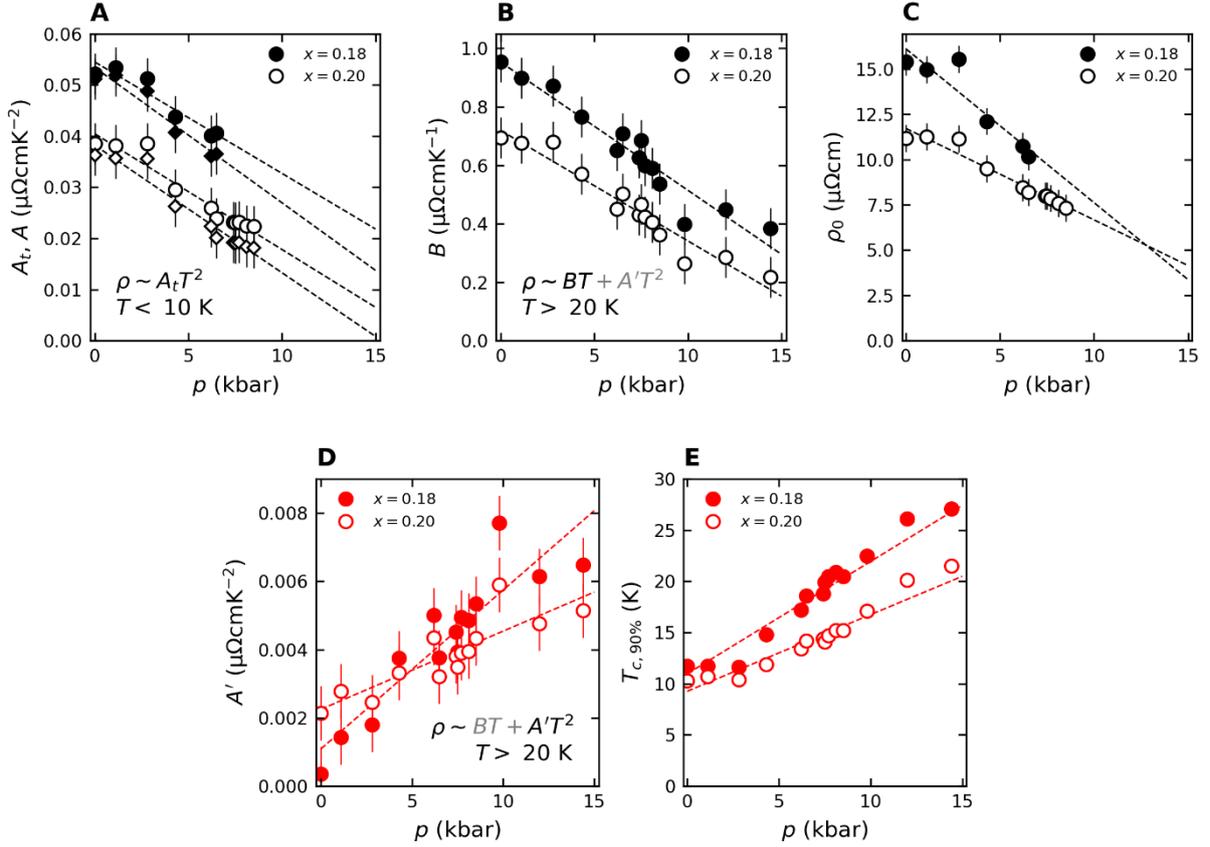

**Fig. 2. Pressure dependence of the resistivity coefficients and superconductivity. A:** Pressure dependence of the low-$T$ $T^2$ coefficient $A_t$ (circles) obtained from linear fits of $d\rho/dT$ below 10 K (black lines in Fig. 1C, D). Also shown are the coefficients $A = A_t - A'$ (diamonds), the component of $A_t$ attributed to electron-electron scattering dressed by critical nematic fluctuations. Dashed lines are linear fits to the data. The strengthening superconductivity prevents $A_t$ (and $A$) from being determined at the highest applied pressures. **B:** Pressure dependence of the $T$-linear coefficient $B$ obtained by fitting $d\rho/dT$ measured between 20 K and 40 K to a straight line. Dashed lines are linear fits to the data. In both panels **A** and **B**, the error bars reflect the sensitivity of each coefficient to details of the fitting procedure (e.g. the precise temperature range used). **C:** Pressure dependence of the residual resistivity $\rho_0$ obtained by extrapolating first of the low-$T$ $\rho(T)$ curves at 35 T to 0 K. The dashed lines extend only up to the pressures at which supercond-ucting fluctuations do not influence $\rho(T)$. **D:** Pressure dependence of the high-$T$ $T^2$ coefficient $A'$ as obtained from straight line fits to $d\rho/dT$ at 35 T. The error bars reflect the sensitivity of $A'$ to the temperature range of each fit. **E:** Pressure dependence of $T_c$ defined as the temperature at which the zero-field resistivity reaches 90% of its value at 35 T. $T_c$ in both samples exhibits an enhancement by a factor of around two.

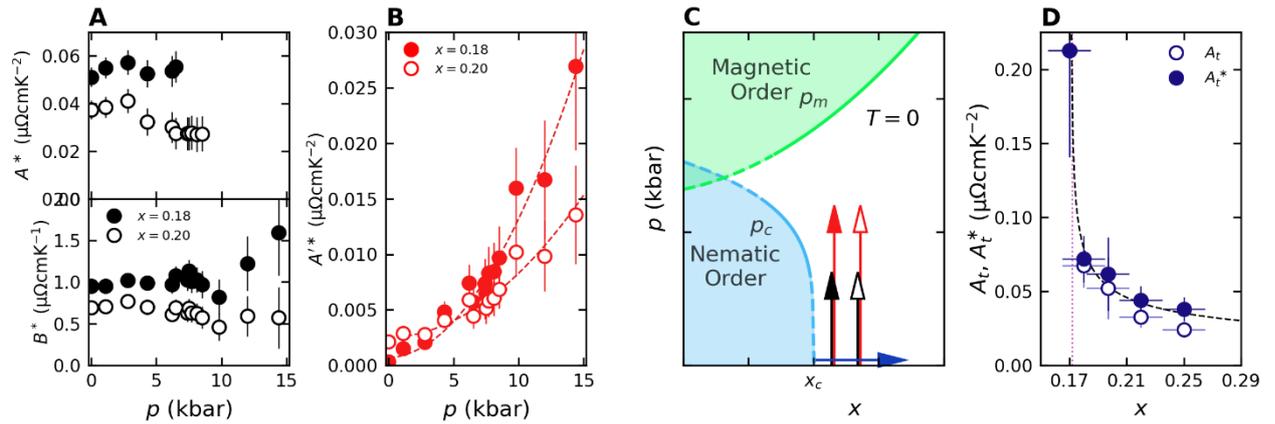

**Fig. 3. Decoupled nematic and magnetic criticality in FeSe$_{1-x}$S$_x$. A:** Renormalized $A^*$ coefficient (upper) and similarly renormalized $B^*$ coefficient (lower) as a function of pressure having rescaled $A$ and $B$ respectively by the pressure-induced change in $\rho_0$. **B:** Similarly renormalized $A'^*$ coefficient as a function of pressure. **C:** Schematic $T = 0$ phase diagram in the $p$-$x$ plane showing the nematic and magnetic phase boundaries. Vertical arrows indicate the pressure-induced approach of the two studied samples to the magnetic QCP and the pressure ranges over which $A$ and $A'^*$ can be determined. The horizontal arrow represents tuning away from the NQCP with increasing $x$, relevant to panel **D**. Near $x = x_c$, the nematic phase boundary is shown as a dashed line to reflect its possible weak first-order nature. **D:** Variation of $A_t$ and $x$ beyond the NQCP near $x_c = \sim 0.17$ (red dotted line). The open (closed) symbols represent, respectively, $A_t$ with (without) rescaling for the relative growth in carrier density with increasing $x$ beyond $x = 0.17$. Vertical (horizontal) error bars reflect scatter in the reported values (uncertainty in $x$), respectively. The dashed line is a guide to the eye. See Supplementary Material for more details of how this plot was conceived.

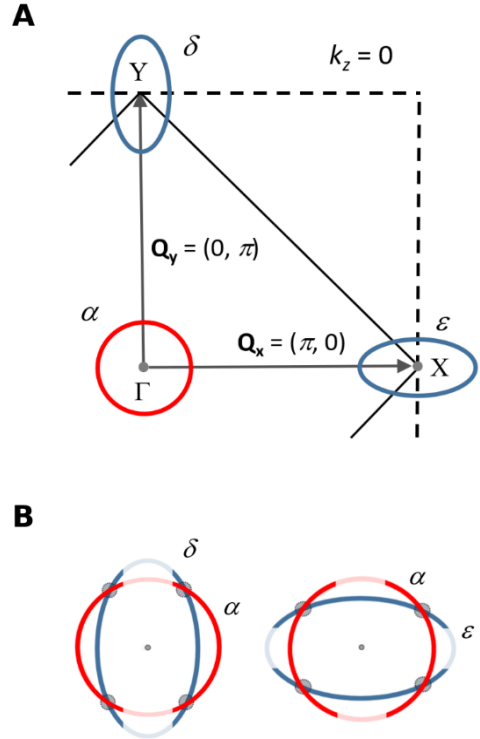

**Fig. 4. Decoupling of the nematic and magnetic interactions and the Fermi surface of FeSe$_{1-x}$S$_x$. A:** Schematic Fermi surface of FeSe$_{1-x}$S$_x$ outside of the nematic phase showing the Γ-centered hole pocket (α) and X,Y-centered electron pockets (ε and δ) at $k_z = 0$. States on different pockets can be connected via finite-**Q** scattering as indicated by the grey arrows. **B:** Schematic illustrating the distinct regions of quasiparticle dressing due to critical magnetic fluctuations (grey circles) arising from translation of the pockets through **Q** = (π, 0), (0, π) and nematic (Pomeranchuk) fluctuations (lighter shaded regions on the electron/hole pockets where the quasiparticle spectral weight is reduced (*48*)).

# Supplementary Text

## Evolution of the $A_t$ coefficients for $x > x_c$ at ambient pressure

The ambient-pressure resistivity in FeSe$_{1-x}$S$_x$ for $x > x_c$ has now been measured in three different laboratories *(34, 36, 37)*. A summary of the $A_t$ coefficients are listed in the second column of Table S1. For $x = 0.17$, the in-plane resistivity $\rho_{ab}(T)$ varies linearly with temperature down to the lowest temperatures studied. Hence, the values plotted in Table S1 are a lower bound for $A_t$ assuming that the crossover to $T^2$ resistivity occurs immediately below the base temperature ($T \sim 3$ K). In order to estimate $A_t$ in this case, we use the gradient of a (fictitious) derivative going from the origin to a value equivalent to the slope of $T$-linear resistivity at the base temperature.

Due to the relative scatter in the data, we list in the first two columns of Table S2 the average values of $A_t$ at each $x$, bar the data for $x = 0.19$ and 0.20, for which the results were binned. These averaged values are then plotted as filled circles in Figure 3D. The error bars associated with these data points reflect the standard deviation (in $A_t$) and the typical spread in $x$ ($= \pm 0.015$) within an individual batch of FeSe$_{1-x}$S$_x$ samples *(12)*.

The dressing of quasiparticles by critical fluctuations manifests both in the enhancement of $\gamma$ – the electronic specific heat coefficient – and $A$ – the coefficient of the $T^2$ resistivity at low $T$. The empirical Kadowaki-Woods ratio $A/\gamma^2$ is found to be of a similar order in many heavy fermion systems *(S1)*, but exhibits a marked variation in magnitude in other correlated metals *(S2)*. This variation is partly due to the fact that $A$ is sensitive not only to $(m^*)^2$, but also to the carrier concentration $n$ *(S2)*. For a quasi-two-dimensional system like FeSe$_{1-x}$S$_x$, $A$ scales as $1/k_F^3$ where $k_F$ is the Fermi wave vector *(S)*. As Se is replaced with S, both pockets increase in size while preserving charge compensation *(40)*. Thus, were $m^*$ to remain constant across the series, $A$ would fall as the effective carrier density increases with $x$. In order to isolate the influence of $m^*$ on $A$, we used the results from the quantum oscillation study reported in Ref. *(40)*. There, typically four frequencies were observed, corresponding to the neck and belly of the two warped cylindrical pockets. Pairing these four frequencies in such a way that charge compensation is preserved and taking the average of each pair, we obtain a single frequency with an effective $k_F(x)$ which, to first order, defines the size of each pocket at every $x$ value studied up to $x = 0.19$. Finally, we extrapolate the linear increase in $n$ between $x = 0.17$ and 0.19 (see Extended Data Fig. 9 in Ref. *(34)*) up to $x = 0.25$ to obtain $n$ for all $x$ values listed in Tables S1 and S2. The resultant $A_t^*$ coefficients listed in the third column of Table S2 and plotted as open circles in Fig. 3D are then obtained from as-measured $A_t$ coefficients using the expression $A_t^* = A_t \times (k_F^3(x)/k_F^3(0.17))$ where $k_F(0.17)$ is the effective, averaged $k_F$ value for the $x = 0.17$ sample.

**Table S1.**

| $x$ | $A_t$ (n$\Omega$cm/K$^2$) | [Ref.] |
|---|---|---|
| 0.17 | 130 | *(34)* |
| 0.17 | 295 | *(34)* |
| 0.18 | 72 | *(36)* |
| 0.18 | 55 | This study |
| 0.18 | 67 | *(34)* |
| 0.19 | 21 | *(36)* |
| 0.20 | 95 | *(34)* |
| 0.20 | 40 | This study |
| 0.22 | 32.5 | *(37)* |
| 0.25 | 22 | *(34)* |
| 0.25 | 29 | *(36)* |
| 0.25 | 22 | *(36)* |

**Table S1**: Measured values of $A_t$, the coefficient of the $T^2$ resistivity, in FeSe$_{1-x}$S$_x$.

**Table S2.**

| $x$ | $A_t$ (nΩcm/K²) | $A_t^*$ (nΩcm/K²) |
|---|---|---|
| 0.17 | 212.5 | 212.5 |
| 0.18 | 65 | 72 |
| 0.197 | 52 | 62 |
| 0.22 | 32.5 | 44 |
| 0.25 | 24 | 38 |

**Table S2**: Columns 1 and 2: Averaged and binned $x$ and $A_t$ values from Table S1 plotted as filled circles in Fig. 3D. Column 3: Renormalized $A_t^*$ values, plotted as open circles in Fig. 3D, obtained by taking into account the change in carrier density using quantum oscillation data from Ref. (*40*). See text for more details.